\documentclass[a4paper,11pt]{article}
\pdfoutput=1 

\usepackage{epsfig}
\usepackage{graphicx}                  
\usepackage{ae}
\usepackage{amsmath}
\usepackage{amssymb}
\usepackage{graphics}
\usepackage{dcolumn}
\usepackage{bm}
\usepackage{lineno}
\usepackage{caption}
\usepackage{ragged2e}

\renewcommand{\thefootnote}{\fnsymbol{footnote}}
\title{\bf Performance study of a bakelite RPC prototype built by new technique of linseed oil coating}
\date{}

\begin{document}

\maketitle
	\flushbottom
\vspace*{-1cm}
\centering

\author{\bf A.~Sen$^a$,}
\let\thefootnote\relax\footnotetext{$^a$Now at Jefferson Lab and Ohio University, Athens, USA}
\author{\bf S.~Mandal,}
\author{\bf S.~Chatterjee$^b$,}
\let\thefootnote\relax\footnotetext{$^b$Now at University of Massachusetts, Amherst, USA} 
\author{\bf S.~Gope,}
\author{\bf S.~Das,}
\author{\bf S. Biswas$^*$}
\let\thefootnote\relax\footnotetext{$^*$Corresponding author. 

\hspace*{0.4cm}E-mail: saikat@jcbose.ac.in }

\vspace*{0.5cm}

	{{Department of Physical Sciences, Bose Institute, EN-80, Sector V, Kolkata-700091, India}

\vspace*{0.5cm}
\centering{\bf Abstract}
\justify

Resistive Plate Chamber (RPC) is one of the most commonly used detectors in high energy physics experiments for triggering and tracking because of its good efficiency ($\textgreater$~90\%) and time resolution ($\sim$~1-2~ns). Generally, bakelite which is one of the most commonly used materials as electrode plates in RPC, sometimes suffers from surface roughness issues. If the surface is not smooth, the probability of micro discharges  and spurious pulses increase, which leads to the deterioration in the performance of the detector. We have developed a new method of linseed oil coating for the bakelite based detectors to avoid the surface roughness issue. The detector is characterised with Tetrafluoroethane (R134a) based gas mixture. The detector is also tested with a high rate of gamma radiation environment in the laboratory for the radiation hardness test. The detailed measurement procedure and test results are presented in this article.






\vspace*{0.25cm}





\section{Introduction}

	Resistive Plate Chamber (RPC), first developed by R. Santonico {\it et al.} \cite{santanico} using bakelite, is a gas filled detector used extensively in high energy physics (HEP) experiments for their high efficiency ($\textgreater$~90\%) and good time resolution ($\sim$~1-2~ns) \cite{Biswas,Park,Bhatt}. HEP experiments such as BaBar \cite{BABAR}, ATLAS \cite{ATLAS}, ALICE \cite{ALICE_muon}, CMS \cite{CMS} use it for the triggering purpose and experiments like ALICE \cite{ALICE_tof}, STAR \cite{STAR} use it as a tracking device. In addition to these, cosmic ray experiments like ARGO-YBJ \cite{ARGO-YBJ}, COVER-PLASTEX \cite{COVER-PLASTEX}, Daya Bay \cite{DAYABAY} also use RPC for the muon detection.
	
	The surface of the bakelite electrode plates normally has some microstructure that can cause a high noise rate and low efficiency \cite{CL09}. In addition to that, this non-uniformity in the electrode surface increases the spark probability and subsequently the leakage current.
	
	To get rid of the surface roughness issue, linseed oil treatment process was already developed in the past \cite{Abbrescia}. Generally after making the gas gap, linseed oil with a thinner (Pentane) solution in some specific ratio is mixed, and the mixture is applied over the bakelite electrode surface as a coating. After the application of the coating, it is completely dried out to cure the electrode plate \cite{hong}. The procedure to apply the linseed oil coating after the gas gap is closed, ensures the cleanliness of the internal electrodes surface.
	
	But in the BaBar experiment, it was first observed that, a too thick and not completely polymerized linseed oil layer gave rise to the formation of stalagmites, and the performance of the RPC was drastically reduced \cite{babar_stalagmite, Anulli_128}.	
	
	After this incident, numerous R\&D on the development of bakelite RPC have been carried out. The problem was finally solved for the linseed oil coated bakelite RPC by using additional thinner (Pentane) with the linseed oil and also using a very thin layer of coating \cite{Atlas,cms}. Sometimes, bakelite RPCs have been fabricated without any coating \cite{Zhang}. Coating of some other oil was also used for bakelite RPCs for low rate experiments \cite{sb_2009}. We have developed a new technique of linseed oil coating to eliminate the curing issue. Usually, the linseed oil coating inside the bakelite RPC is done after making the gas gap. In this procedure, the oil coating is done before fabricating the detector to check visually whether the curing is properly done, or any uncured droplet of linseed oil is present. The details of the fabrication procedure and the first performance studies are presented elsewhere \cite{sen_2022}. The fabrication process is also briefly described in Sec.~\ref{const}.
	

	
	\begin{center}

		\begin{table}[htb!]
								\caption{Summary of previous experimental results} \label{table1}

			\begin{center}
				
				\resizebox{\columnwidth}{!}{
					\begin{tabular}{|c|c|c|c|} \hline
					{\bf Gas composition} & {\bf Efficiency$^*$} & {\bf Noise rate} & {\bf Leakage current }\\ 
					{\bf } & {\bf (\%)} & {\bf  at} & {\bf per unit area}\\
					{\bf } & {\bf  } & {\bf  plateau} & {\bf @ 10~kV}\\

					{\bf } & {\bf  } & {\bf  (Hz/cm$^2$)} & {\bf ($\mu$A/m$^2$)}\\ \hline

				        100\% Tetrafluoroethane (R134a)  &  \multicolumn{2}{c|}{ @ -~1.5 mV threshold}    & 260.6 \\

				     (100\% C$_2$H$_2$F$_4$)    &  95$\pm$1  & 500   & \\ 
					
			               &   from 9.4 kV onwards  &  & \\ 
			               			               &--------------------------&------------& \\ 

			                 &  \multicolumn{2}{c|}{ @ -~2 mV threshold}    &  \\ 
			                 
			                     &  85$\pm$5   & 200   & \\ 
					
			               &   from 10.1 kV onwards  &  & \\ 
			                 
			                 \hline

				C$_2$H$_2$F$_4$ / i-C$_4$H$_{10}$ 	  &  \multicolumn{2}{c|}{ @ -~2 mV threshold}    & 61.7 \\

				90/10         &  95$\pm$2   & 120   & \\ 
					
			               &   from 10 kV onwards  &  & \\ 
			               			               &--------------------------&------------& \\ 

			                 &  \multicolumn{2}{c|}{ @ -~2.5 mV threshold}    &  \\ 
			                 
			                     &  95$\pm$2   & 80   & \\ 
					
			               &   from 10 kV onwards  &  & \\ 
			                 
			                 \hline

					\end{tabular}
				}
			\end{center}
			{$^*$ Efficiency is defined in Section~~\ref{exp_set_up}.} 
		\end{table}
		
	\end{center}

	The detector is initially tested with 100\% Tetrafluoroethane (R134a) (C$_2$H$_2$F$_4$) gas, and an efficiency $\textgreater$~90\% with higher noise rate compared to the conventional linseed oil-coated detectors is obtained \cite{sen_2022}. The detector is further characterised with conventional mixed gas $i.e.$ 90\% C$_2$H$_2$F$_4$ and 10\% iso-butane (i-C$_4$H$_{10}$). An efficiency greater than 90\% is also found with  this mixture but with a lower noise rate \cite{sen_2022_2}. The leakage current is also found to be lower with the mixed gas compared to that with the 100\% C$_2$H$_2$F$_4$. At a glance, the previously obtained experimental results for two different gas compositions are summarised in Table~\ref{table1} \cite{sen_2022, sen_2022_2}. A typical gas flow rate of $\sim$~2 cc/min equivalent to $\sim$~20 gap volume changes per day is maintained during the test.

	Radiation hardness is one of the important factors for detectors in heavy-ion physics experiments. That is why the efficiency and noise rate of the chamber is measured in the laboratory with a gamma-ray background. In this article, the results of time resolution measurement and long-term stability test are presented along with the result of the radiation hardness test.
	
	\section{Measurement of surface resistivity}
	
To evaluate the effect of the linseed oil coating on the surface roughness, the latter is indirectly assessed by measuring the surface resistivity. It is found in literature that there are several models as well as experimental results where it is mentioned that surface resistivity increases due to surface roughness \cite{PZ2012, SPM1949, SG2017, ME1971}. As the surface resistivity of bakelite surface is quite high, it can not be measured by only using a multimeter. It is done by using an alternative simple circuit.

	\begin{figure}[htb!]
		\centering{
			\includegraphics[scale=0.38]{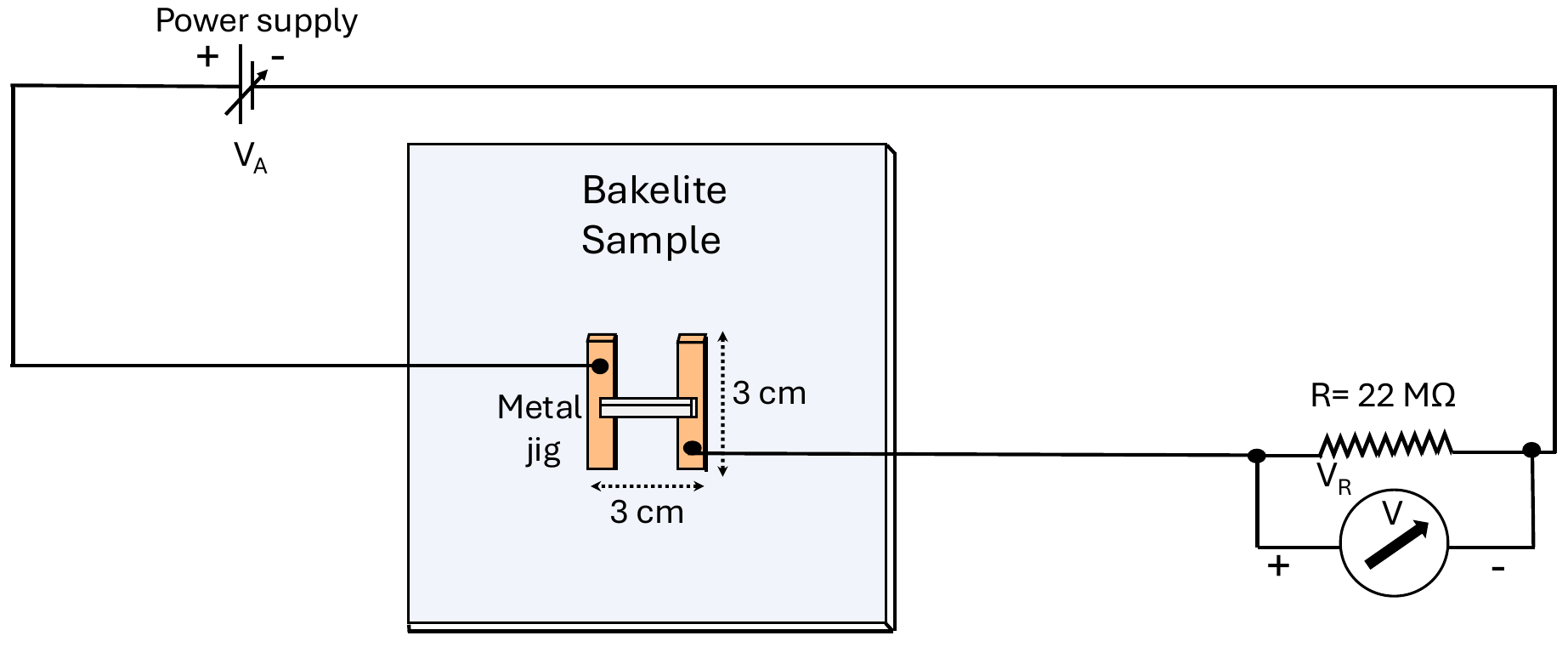}
		}
		\caption{Schematic diagram of the surface resistivity measurement circuit.}
		\label{surfacediagram}
	\end{figure}

A bakelite plate of same type by which the RPC prototype is built is taken and one side of it is coated with linseed oil exactly in the same way it is done for the RPC electrode. After curing, the surface resistivity is measured using the circuit shown in Figure~\ref{surfacediagram}. The series circuit contains a power supply and a high resistor of value 22~M$\Omega$ (R). The voltage across the 22~M$\Omega$ resistor is measured by a Keithley multimeter.  An aluminium jig of 'H' shape is taken with sides of length 3~cm and having a separation between the bars of 3~cm. The two arms of the jig were separated by a G-10 bar. The sides of the aluminium jig are covered with copper tape for better smoothness, better connectivity with the surface and we can solder the connecting wires on the bars as shown in the figure. The voltage, V$_A$ is applied and the voltage across the 22~M$\Omega$, V$_R$ is measured using a multimeter, which is then used to obtain the current through the entire circuit. From this current the total resistance of the circuit is calculated. Finally the surface resistivity of the bakelite sample is calculated in the unit of $\Omega$/$\Box$ using the formula
	\begin{equation}
	\begin{split}
		R_{surface} = 	(\frac{V_A}{V_R} - 1)~\times~R 
		\end{split}
\end{equation}

Using this method the surface resistivity is measured on both the linseed oil coated and uncoated surfaces of the plate at 12 different places, moving the jig. The distribution of the surface resistivity measured for two surfaces are shown in Figure~\ref{surfaceres_result}. For the uncoated surface the average surface resistivity is found to be 1.37~$\times$~10$^{11}$~$\Omega$/$\Box$ with a standard deviation of 7.48~$\times$~10$^{10}$~$\Omega$/$\Box$ and for the linseed oil coated surface it is found to be 8.22~$\times$~10$^{10}$~$\Omega$/$\Box$ with a standard deviation of 4.29~$\times$~10$^{10}$~$\Omega$/$\Box$. It can be seen that after linseed oil coating the surface resistivity decreased referring that the surface roughness also decreased. From the standard deviation it is also observed that the uniformity in surface resistivity is better in the linseed oil coated surface.
	
	
	\begin{figure}[htb!]
		\centering{
			\includegraphics[scale=0.3]{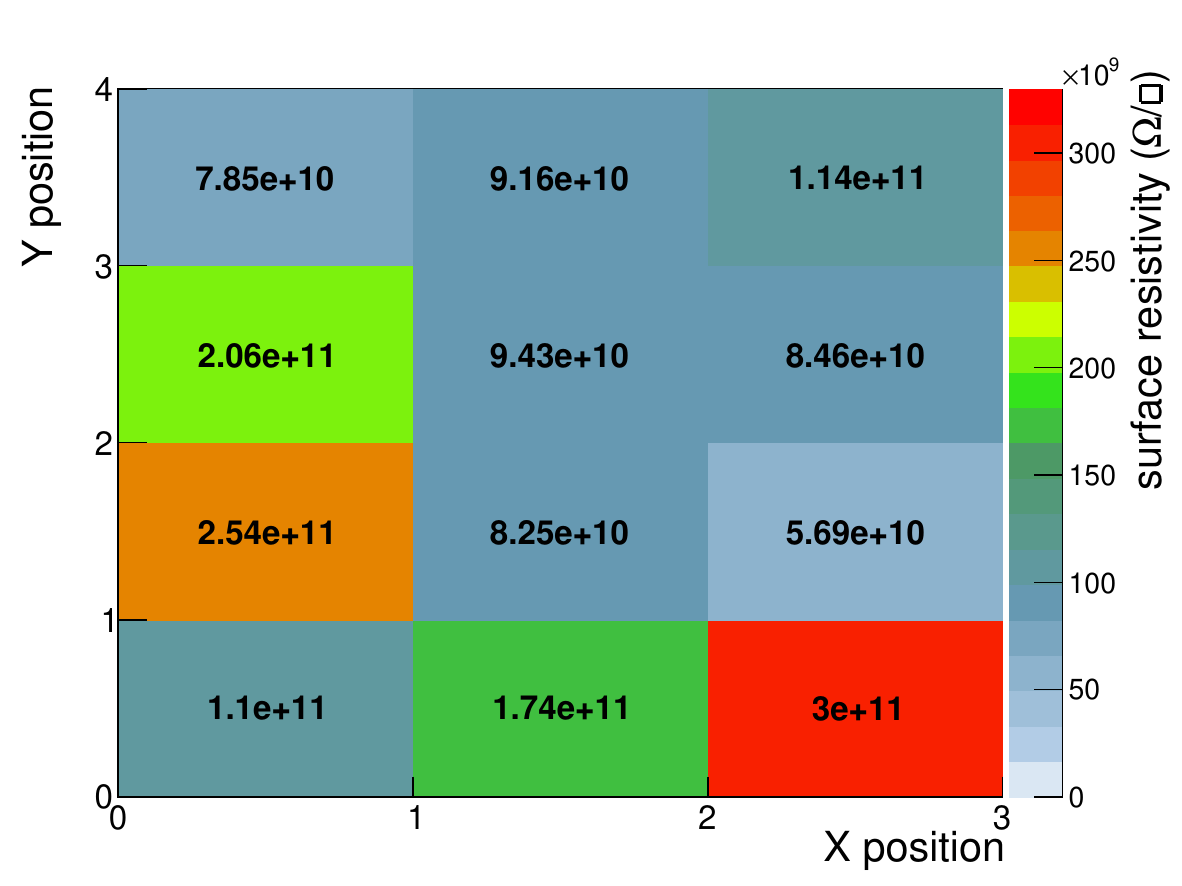}
			\includegraphics[scale=0.3]{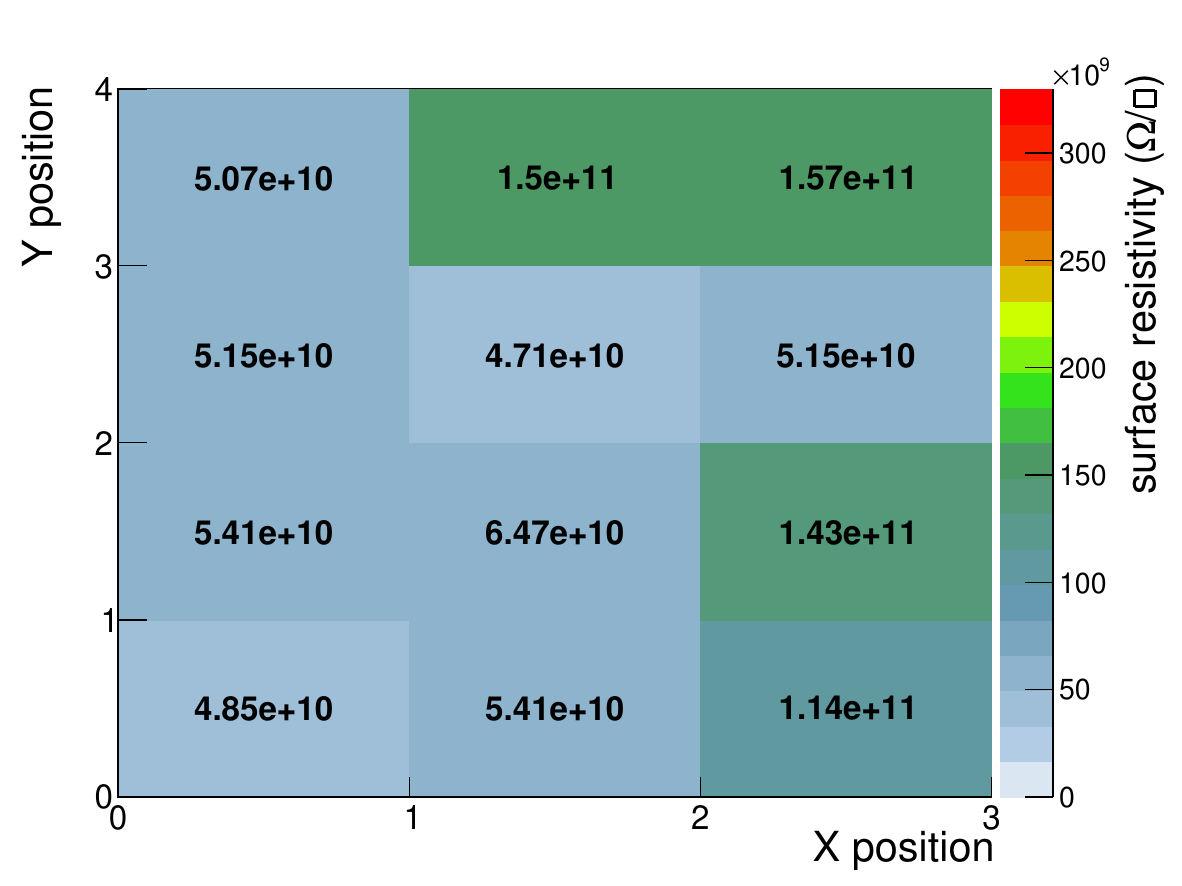}

		}
		\caption{Surface resistivity distribution of uncoated (left) and linseed oil coated (right) surfaces.}
		\label{surfaceres_result}
	\end{figure}
	

	\section{Construction of the chamber}\label{const}
	
	The electrode plates for the present prototype are made of commercially available bakelite high pressure paper laminates of dimensions 27~cm~$\times$~27~cm and thickness of 2~mm each having bulk resistivity $\sim$~3~$\times$~10$^{10}$~$\Omega$~cm at 25$^{\circ}$C. Before making the gas gap, a thin layer of linseed oil coating is applied to the inner surfaces of the electrode plates to make the surface smooth. About 2~g linseed oil is applied over the 27~cm~$\times$~27~cm area of each electrode plate. Based on the specific gravity (0.930 at 15.5 $^\circ$C) of the fluid, the coating thickness is estimated to be $\sim$~30~$\mu$m \cite{sen_2022}. The linseed oil is cured for about 15 days and checked visually whether the oil is dried out completely. Four edge spacers of dimensions 27~cm~$\times$~1~cm and one button spacer of diameter 1~cm are used for making the uniform gas gap of 2~mm. Two gas inlets and outlet nozzles are placed along with the edge spacers at diagonal corners. All the spacers and gas nozzles are made of polycarbonate (bulk resistivity $\sim$~10$^{15}$~$\Omega$~cm) and have a thickness of 2~mm. The drawing of the cross-section of the detector is shown in Figure~\ref{cross_section}. The average surface resistivity of the two outer surfaces having  graphite layer are measured to be $\sim$~510~k$\Omega$/$\Box$ and $\sim$~540~k$\Omega$/$\Box$ respectively. High voltage (HV) of equal and opposite polarities are applied on the diagonally opposite corners of the two outer surfaces of the detector. The oil coating on the bakelite plates, making of gas gap and gluing; are done inside a laminar flow table to ensure the cleanliness.
	
	\begin{figure}[htb!]
		\centering{
			\includegraphics[scale=0.4]{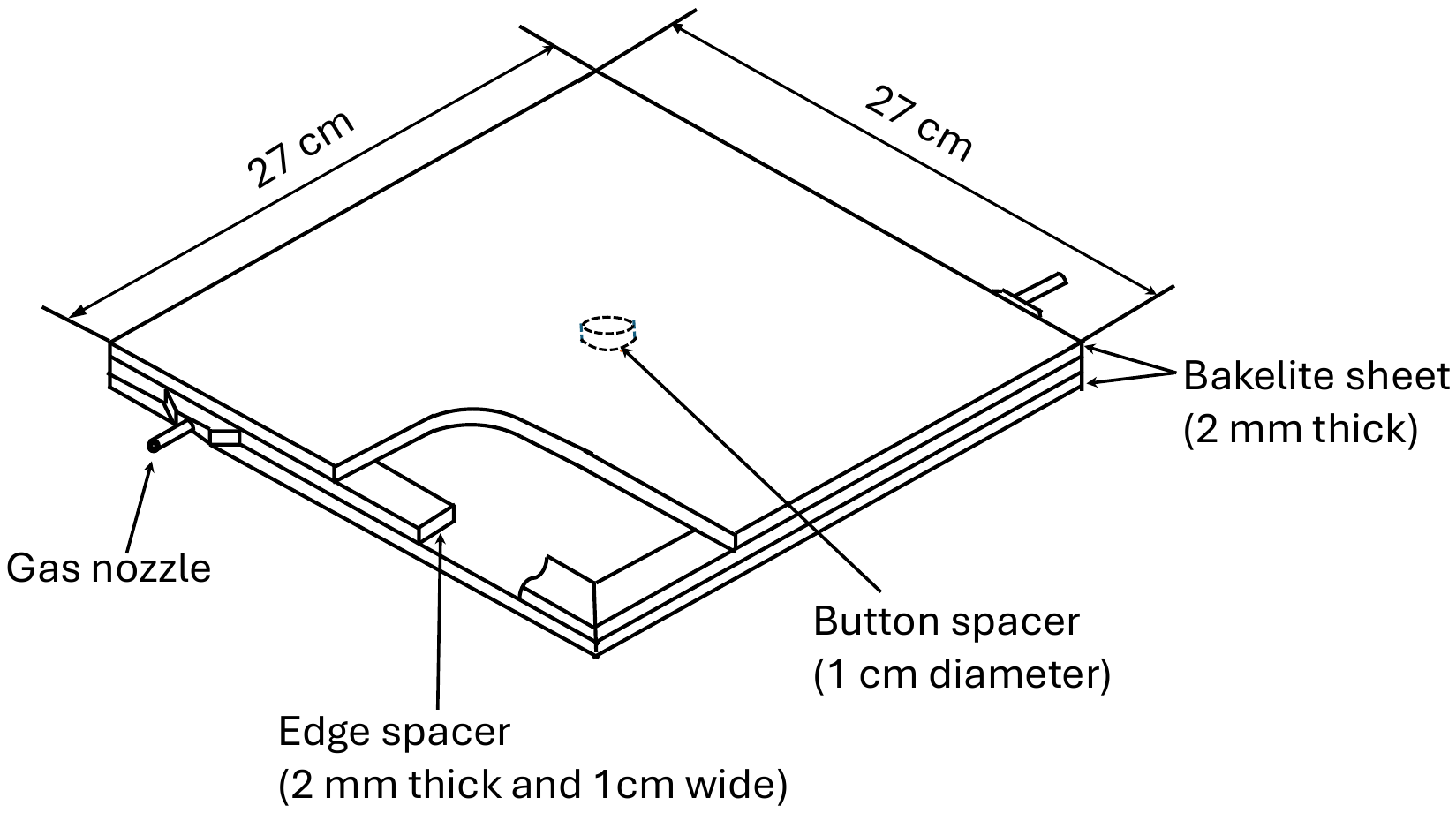}
		}
		\caption{Schematic diagram of the prototype resistive plate chamber.}
		\label{cross_section}
	\end{figure}
	
	Signals are collected using 2.5~mm wide copper strips. The separation of 2~mm is kept between two consecutive strips. The signals from the strips are collected through RG-174/U coaxial cables. More details on the fabrication procedure can be found in \cite{sen_2022}.

	\section{Experimental set-up}\label{exp_set_up}
	
	In this work, the efficiency, noise rate, time resolution, charge sharing,  long-term stability of the prototype are measured using cosmic rays. The efficiency of the chamber is also measured in a high gamma ray background in the laboratory using $^{137}$Cs (662 keV gamma) source of activity 13.6 GBq.
		
	To generate the cosmic ray trigger, an array of three plastic scintillation detectors of thickness 1~cm is used. On top of the RPC, a scintillator of dimension 10~cm~$\times$~10~cm (SC1) and a finger scintillator of dimension 10~cm~$\times$~2~cm (SC2) are placed, and at the bottom, the paddle scintillator of dimension 20~cm~$\times$~20~cm (SC3) is placed as shown in Figure~\ref{cosmic_circuit}. The separation between the SC1 and SC2 is kept 5~cm whereas the distance between SC2 and RPC and RPC and SC3 are kept at 3~cm and 4~cm respectively. +1550~V is applied to each PMT for the operation of the scintillators. -15~mV threshold is applied to the Leading Edge Discriminator (LED) for each scintillation detector to get rid of noise. Discriminated signals from three scintillation detectors are taken in coincidence to generate the 3-fold trigger (master trigger).
	
	\begin{figure}[htb!]
		\centering{
			\includegraphics[scale=0.44]{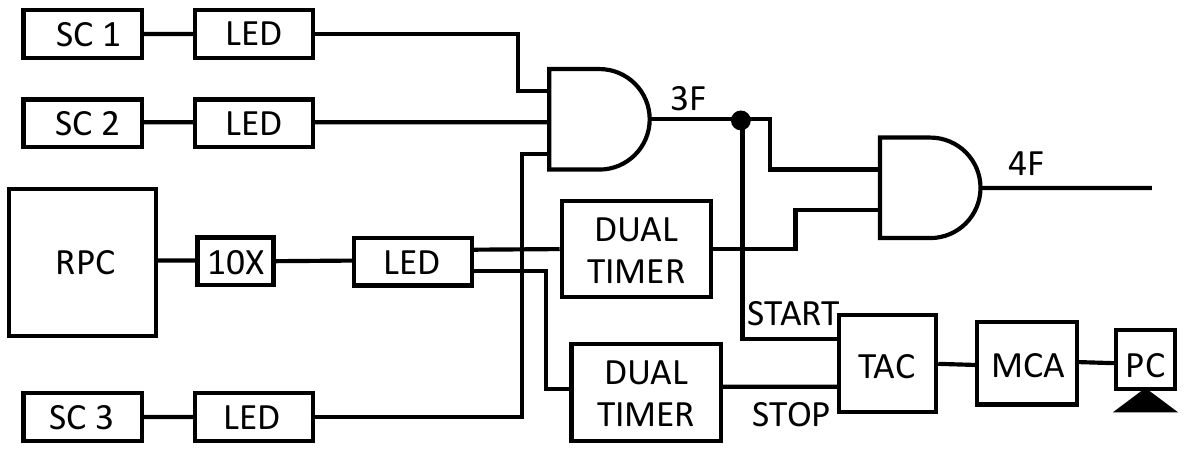}
		}
		\caption{Block diagram of the cosmic ray test setup for the measurement of efficiency and time resolution. SC1, SC2 and SC3 are the plastic scintillators of dimensions 10~cm~$\times$~10~cm, 10~cm~$\times$~2~cm and 20~cm~$\times$~20~cm respectively. The distance between SC1 and SC2, SC2 and RPC, and RPC and SC3 are kept at 5~cm, 3~cm and 4~cm respectively. 10X, LED, TAC, MCA, PC are the 10X fast amplifier, Leading Edge Discriminator, Time to Amplitude Converter, Multi-Channel Analyser and Personal Computer respectively.}
		\label{cosmic_circuit}
	\end{figure}

	\begin{figure}[htb!]
		\centering{
			\includegraphics[scale=0.44]{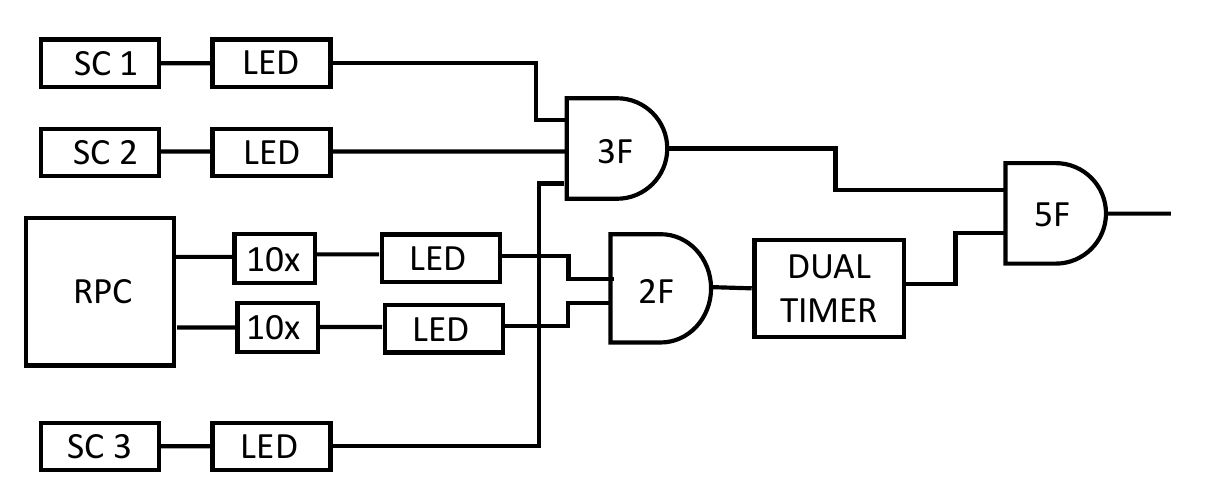}
		}
		\caption{Block diagram of the setup for charge sharing measurement.}
		\label{circuit_charge_sharing}
	\end{figure}
	
	The signal from the RPC pickup strip is fed to a 10X fast amplifier and the amplifier signal goes to the LED. One output from the LED goes to the scalar counter for the singles count or the noise count of the detector. For a finite time interval the counts from one single RPC strip is stored in the scalar. When the number is divided by the time interval and the area of the strip the single count rate is found in unit of Hz/cm$^{2}$. The other output from the LED is fed to a Dual Timer where the discriminated RPC signal is stretched to 500~ns to avoid any double or multiple counting of the signals and also to apply proper delay ($\sim$~250~ns) to match the RPC signal with the 3-fold trigger. The output of the dual timer is finally fed to the logic module to get the 4-fold signal generated in coincidence with the 3-fold trigger. The ratio of the 4-fold signal to the 3-fold trigger signal is defined as the efficiency. The area of coincidence window of the cosmic ray test-bench is 10~cm~$\times$~2~cm.
	
	
	
	To measure the time resolution of the detector, the 3-fold trigger signal is used as the START signal of the Time to Amplitude Converter (TAC) and the dual timer output (as shown in Figure~\ref{cosmic_circuit}) is used as the STOP input signal of the TAC. The output of the TAC is further fed to the Multi-Channel Analyser (MCA), and MCA is connected to the Personal Computer (PC) to store the timing spectra. 100\% C$_2$H$_2$F$_4$ gas is used as the active medium during the study of the timing properties. The temperature and the humidity are also recorded during the entire measurements using a data logger, built in-house \cite{sahu}.

	For measurement of charge sharing, signals from two consecutive strips are taken in coincidence as shown in the Figure~\ref{circuit_charge_sharing}. The amplified and discriminated RPC signals from two consecutive strips are first sent to the coincidence logic and the logic output (2F) is fed to the dual timer for the proper delay matching. Dual timer output is then taken in coincidence with the trigger generated by 3-fold scintillator array to make 5F. The ratio of the 5F count to the 3F count is defined as the charge sharing.

	For the radiation hardness measurement, a strong $^{137}$Cs (662~keV gamma) source of activity 13.6~GBq is used.

		\section{Result}
		
	The detector is first tested with the 100\%	C$_2$H$_2$F$_4$ gas using cosmic ray. Efficiency plateau  $\sim$~95\% from 9.4 kV onwards and $\sim$~85\% from 10.1 kV onwards are obtained for -1.5~mV and -2~mV discriminator threshold\footnote{The signals from the RPC pickup strip are fed to 10X fast amplifier and the amplified signal goes to the LED. So, the threshold -15~mV -20~mV etc. at LED after 10X amplifier means the threshold to the RPC signal is -1.5~mV and -2~mV respectively. In this paper, threshold to the RPC signal means threshold before 10X fast amplifier.} settings respectively. The noise rate measured is very high for the lower threshold with a maximum value of $\sim$~500~Hz/cm$^{2}$ for the prototype \cite{sen_2022}.
	
	The detector is further tested with a gas mixture of C$_2$H$_2$F$_4$ and i-C$_4$H$_{10}$ in the 90/10 volume ratio. Isobutane has a high UV absorption coefficient, and it prevents the formation of secondary discharges due to photoelectrons. The performance is even better with the application of an additional quencher. Both the current and noise rate are very low for this gas mixture compared to that with the 100\% C$_2$H$_2$F$_4$ used for the same detector. An efficiency greater than 90\% is achieved from 10~kV onwards for both -~2~mV and -~2.5~mV threshold settings. The maximum noise rates are found to be 120~Hz/cm$^{2}$ and 80~Hz/cm$^{2}$ for the -~2~mV and -~2.5~mV thresholds, respectively. The detailed I-V characteristics, efficiency, and noise rate results are elaborated in the article \cite{sen_2022_2}. It is to be mentioned here that the noise rates obtained here are one or some time two orders of magnitude higher compared to that obtained in Ref.~\cite{Abbrescia}.
		
	To measure the timing properties, the RPC signal is stretched to 500~ns. Full scale of the TAC is set to 100~ns. The START signal of the TAC is taken from the 3-fold scintillator trigger and the STOP signal is taken from the RPC. The typical time spectrum at 10.2~kV for the RPC with 100\% C$_2$H$_2$F$_4$ gas is shown in Figure~\ref{time_spectrum}. The distribution of the time difference between the master trigger and the RPC signal is plotted. The $RMS_{tot}$ of the distribution is found out and subtracting the contribution of the scintillators ($RMS_{SC}$) in quadrature the intrinsic time resolution ($RMS$) of the RPC detector is calculated using the formula;  

	\begin{equation}
	\begin{split}
		RMS = \sqrt{RMS_{tot}^2~-~RMS_{SC}^{2}}
			\end{split}
\end{equation}

Before measuring the time resolution of the RPC, the intrinsic time resolutions of all the scintillators are measured by different combination, i.e. taking the START signal of the TAC from one scintillator and STOP signal from other scintillator and so on. At 10.2~kV the $RMS_{tot}$ is found to be 1.69~$\pm$~0.03~ns as shown in Figure~\ref{time_spectrum}. The measurement is repeated for five different voltage settings.

	
	\begin{figure}[htb!]
		\centering{
			\includegraphics[scale=0.44]{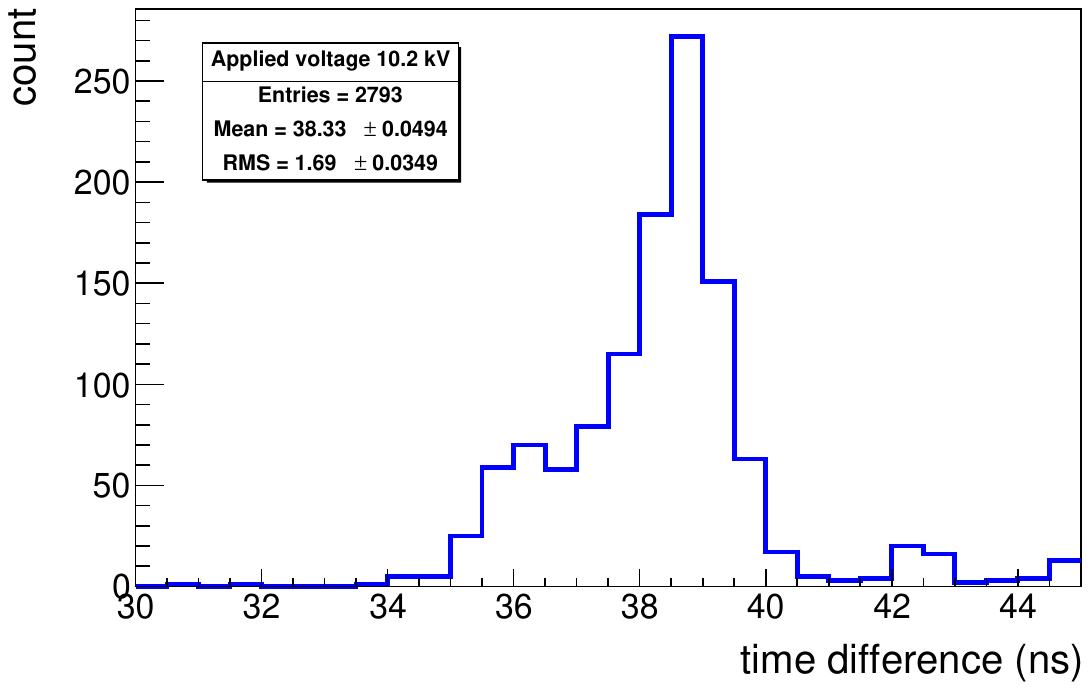}
		}
		\caption{Time spectrum of RPC at an applied voltage difference of 10.2 kV across the gas gap.}
		\label{time_spectrum}
	\end{figure}
	
	
	The intrinsic time resolution ($RMS$) of the RPC as a function of the applied voltage is shown in Figure~\ref{sig}. With the increase of applied voltage RMS improves. At 10.6~kV, the optimum time resolution (RMS) is found to be $\sim$~1.03~$\pm$~0.03~ns whereas the RPC time resolution (RMS) at 10.2~kV is $\sim$~1.65~$\pm$~0.03~ns. It is to be mentioned here that previously for a prototype made with the same material but without oil coating, the resolution was found to be $\sim$~1.2~ns ($\sigma$) at 10.2~kV \cite{sen_2020}.
	
	
	\begin{figure}[htb!]
		\centering{
			\includegraphics[scale=0.44]{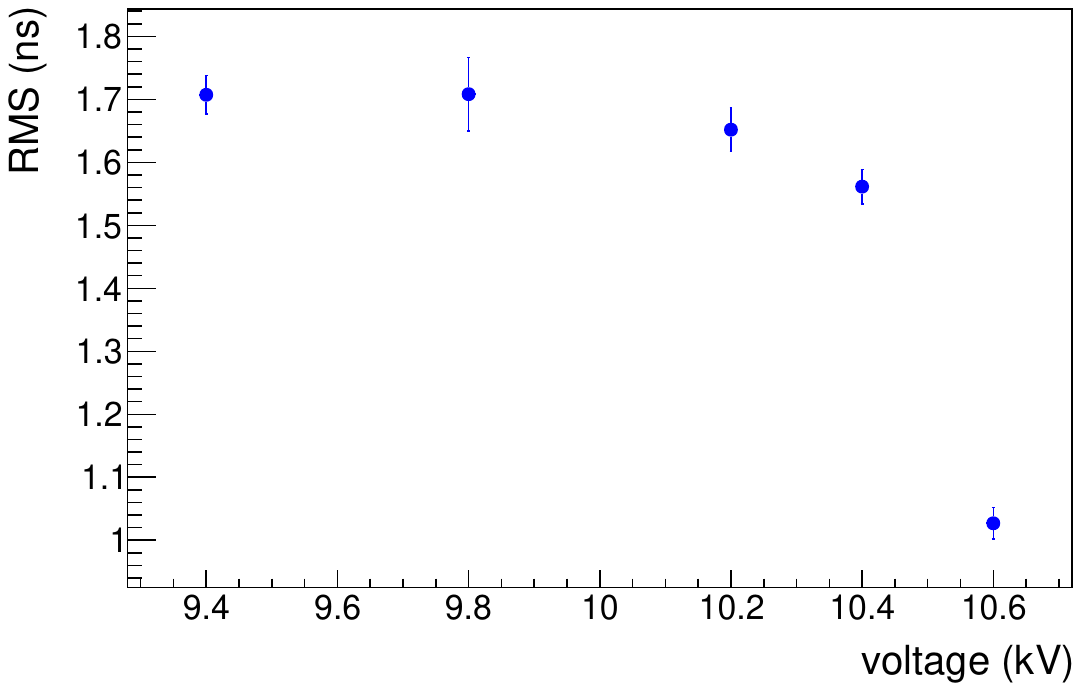}
		}
		\caption{Time resolution ($RMS$) of the RPC as a function of applied voltage.}
		\label{sig}
	\end{figure}
	
	
Charge sharing is measured between the two consecutive readout strips varying the applied voltage with the C$_{2}$H$_{2}$F$_{4}$ and i-C$_{4}$H$_{10}$ gas mixture. As mentioned earlier the charge sharing is defined as the ratio of the coincidence count of two consecutive readout strips taken in coincidence with the 3F scintillator trigger (5F) to the trigger count (3F). Actually, for this measurement the finger scintillator of the trigger is placed just above one single strip and from which the efficiency is also measured for reference.
	  
The charge sharing between two consecutive strips as a function of applied voltage in shown in Figure~\ref{charge_shr}. In the same plot the efficiency as a function of voltage measured on the same day is also shown for reference. One can see in Figure~\ref{charge_shr} that the shared charge is about $\sim$~50\%, where the efficiency measured from a single strip is  $\sim$~82\%. For a variation of voltage between 5~-~9~kV, the charge sharing remains at the level of 40-50\%. Although it has been referred as the measurement of charge sharing, the crosstalk between the two strips is not eliminated for this particular measurement.

	
	\begin{figure}[htb!]
		\centering{
			\includegraphics[scale=0.44]{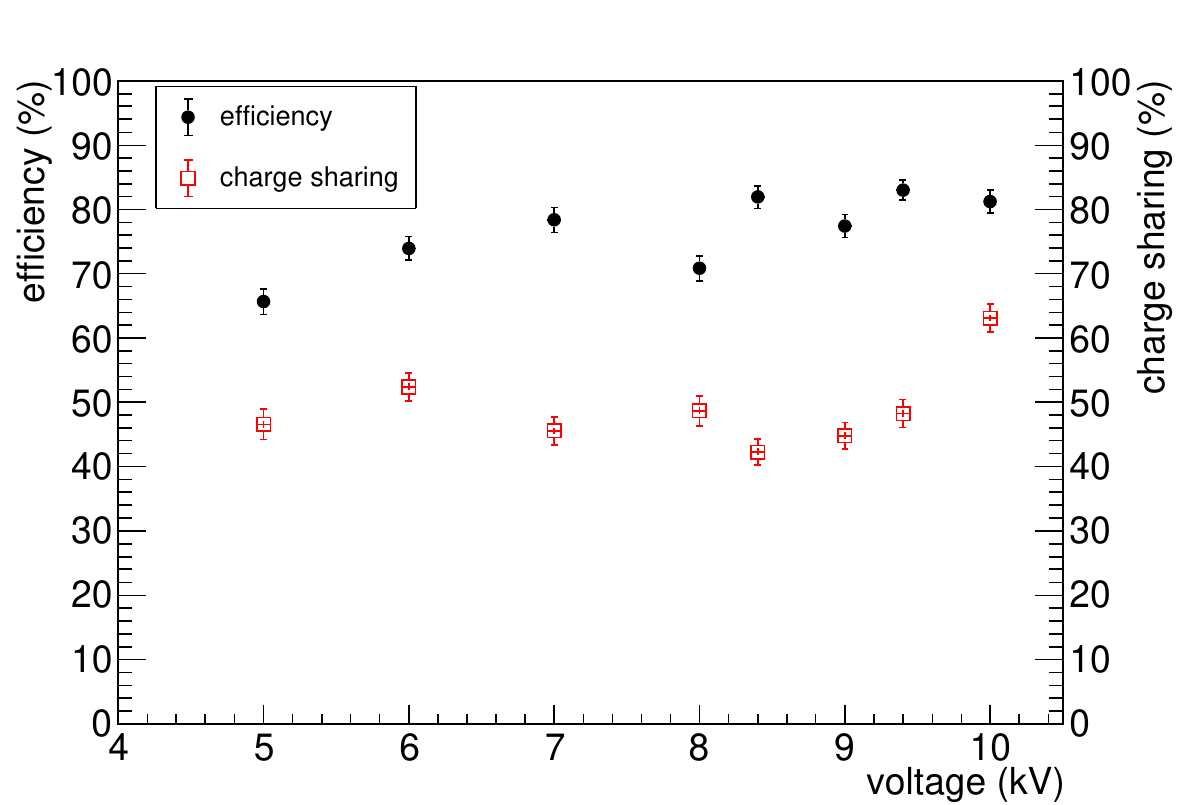}
		}
		\caption{Variation of efficiency and shared charge between two consecutive strips with the voltage.}
		\label{charge_shr}
	\end{figure}
	

Finally, the prototype is tested in the high intensity gamma ray environment. A $^{137}$Cs source of activity 13.6 GBq is used for this measurement. 662~keV photons are emitted from the source with a measured flux of $\sim$~46~kHz/cm$^{2}$. As shown in Figure~\ref{rad_setup} the source is placed on the top of the top scintillator paddle. The efficiency is measured for three different voltage settings with and without the gamma source. The measurement is performed with the C$_{2}$H$_{2}$F$_{4}$ and i-C$_{4}$H$_{10}$ gas mixture in the 90/10 volume ratio. The obtained efficiency value as function of voltage across the gas gap is shown in Figure~\ref{rad_result} in absence and presence of the gamma source. It is observed that the efficiency with the source decreased by only 1~\% from the efficiency value without the source. 
	
	
	\begin{figure}[htb!]
		\centering{
			\includegraphics[scale=0.18]{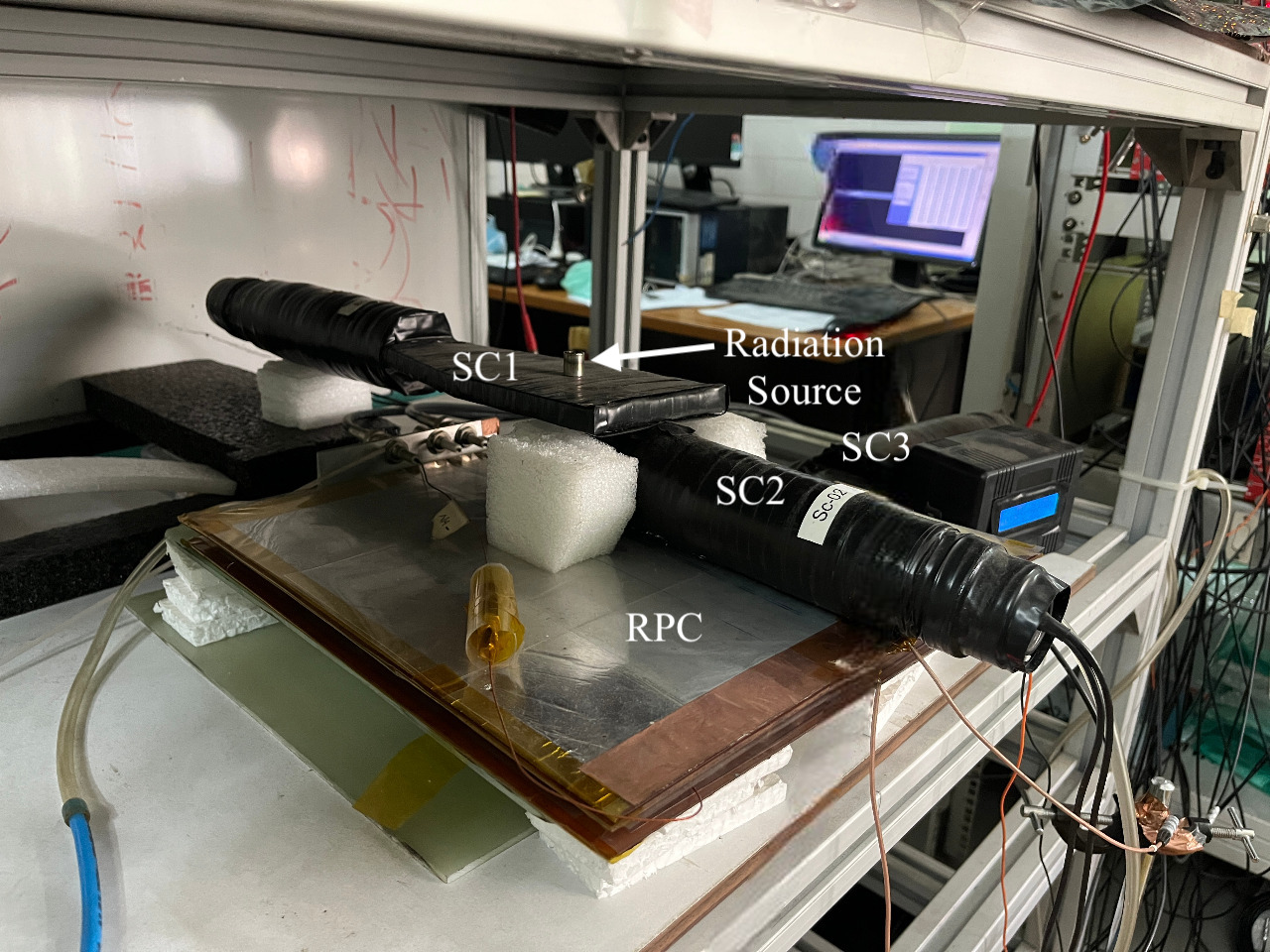}
		}
		\caption{Setup to measure the efficiency in presence of high intensity gamma ray flux.}
		\label{rad_setup}
	\end{figure}
	

	
	\begin{figure}[htb!]
		\centering{
			\includegraphics[scale=0.5]{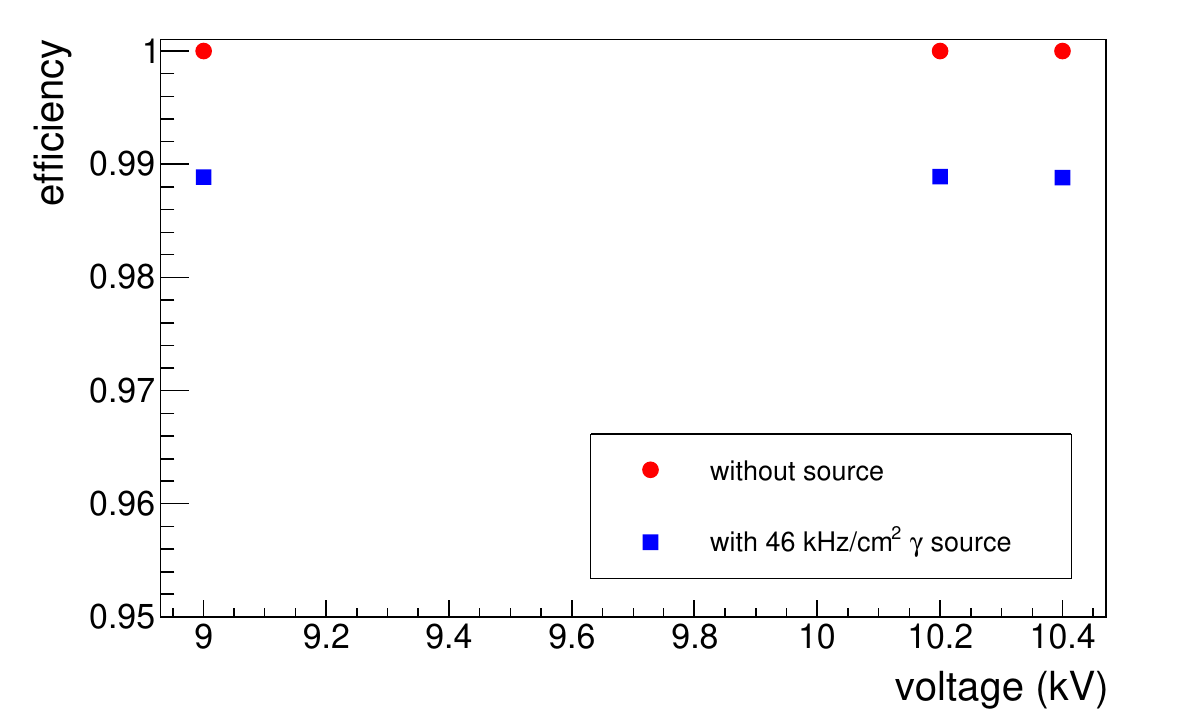}
		}
		\caption{Efficiency as a function of voltage across the gas gap in the presence and absence of high intensity gamma ray flux.}
		\label{rad_result}
	\end{figure}
	

The detector is also tested for long-term with cosmic rays. During the long-term operation the detector is operated with two types of gas mixtures. One is 100\% C$_2$H$_2$F$_4$ gas and another is  C$_2$H$_2$F$_4$ and i-C$_4$H$_{10}$ mixture in the 90/10 volume ratio. Initially, the detector is operated with a mixture of C$_{2}$H$_{2}$F$_{4}$ and i-C$_{4}$H$_{10}$ and then it is continuously tested with 100\% C$_2$H$_2$F$_4$ for next few days. After that it is again tested with the mixed gas.

	
	\begin{figure}[htb!]
		\centering{
			\includegraphics[scale=0.43]{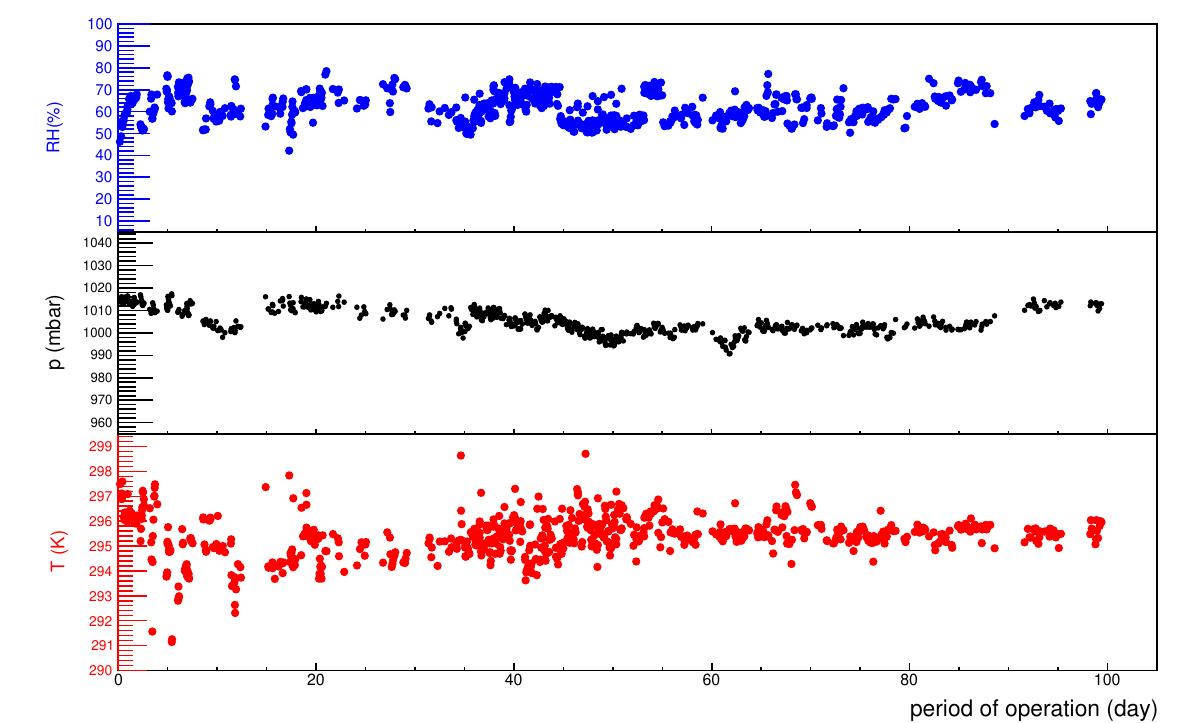}
		}
		\caption{Temperature in Kelvin, pressure in mbar and \% relative humidity as a function of period of operation.}
		\label{eff_parameter}
	\end{figure}
	
	
	
	\begin{figure}[htb!]
		\centering{
			\includegraphics[scale=0.43]{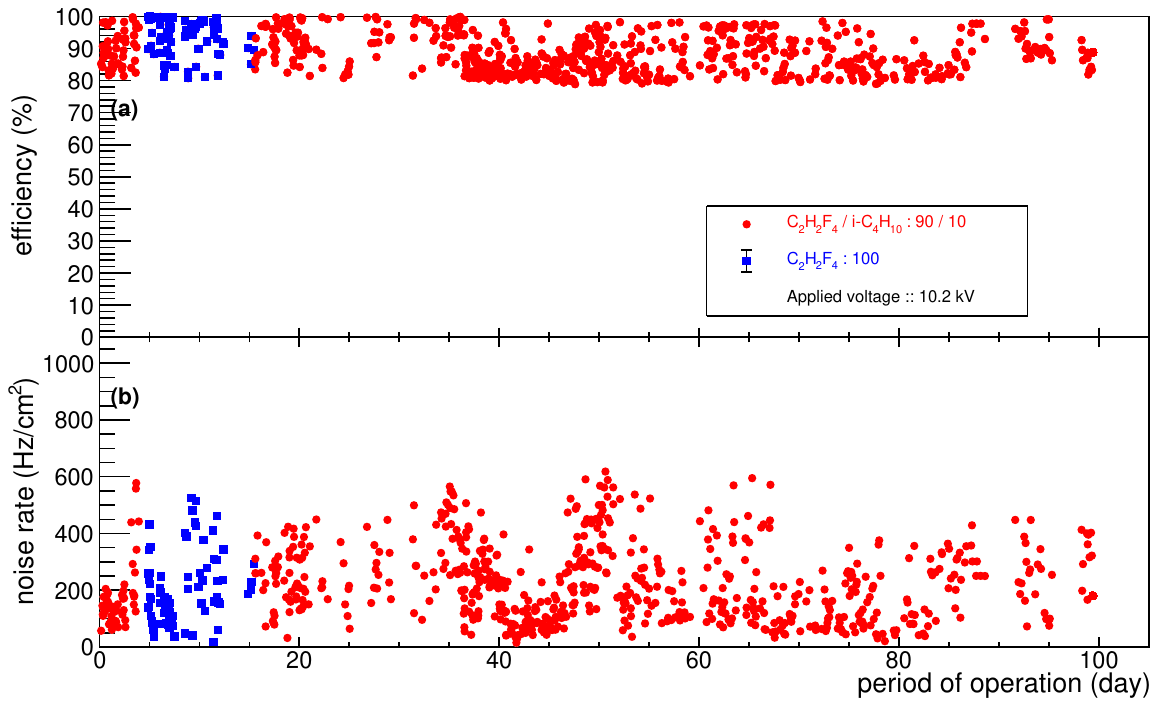}
		}
		\caption{(a) Efficiency and (b) noise rate of the detector as a function of period of operation for two different gas compositions at applied voltage of 10.2~kV. For some data points the error bars are smaller than the size of the markers.}
		\label{eff_noise}
	\end{figure}
	


\begin{figure}[htb!]
	\centering{
		\includegraphics[scale=0.42]{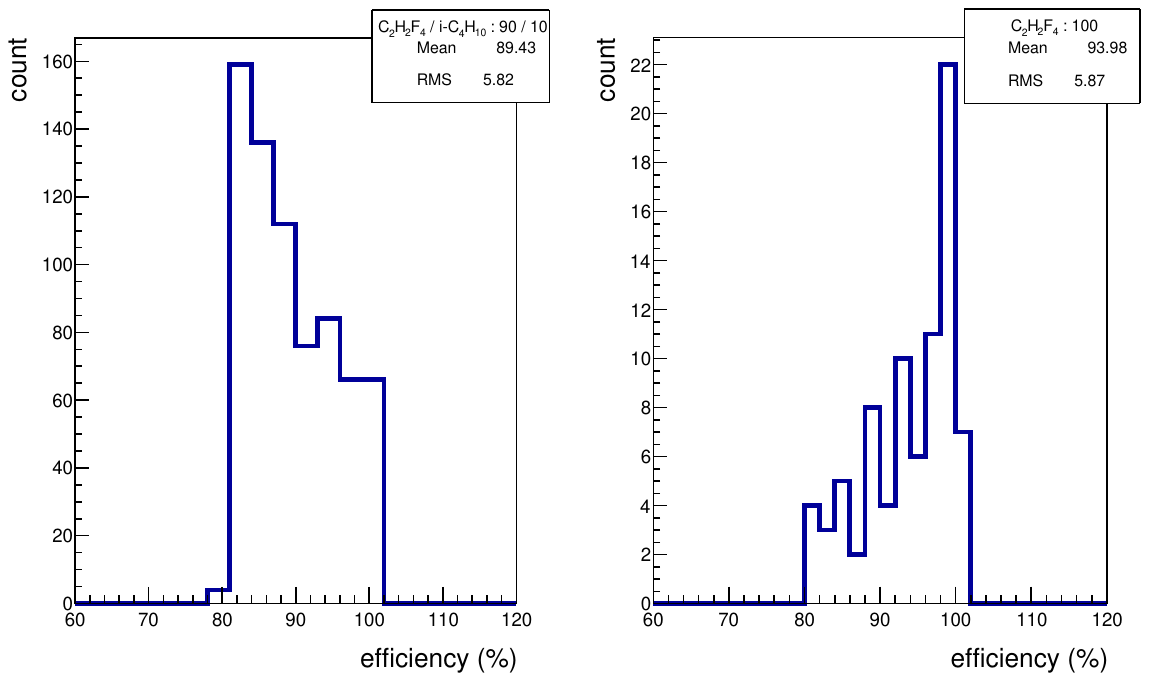}
	}
	\caption{Distribution of the efficiency values of long-term measurements at applied voltage of 10.2~kV.}
	\label{eff_distribution_long_term}
\end{figure}



\begin{figure}[htb!]
	\centering{
		\includegraphics[scale=0.42]{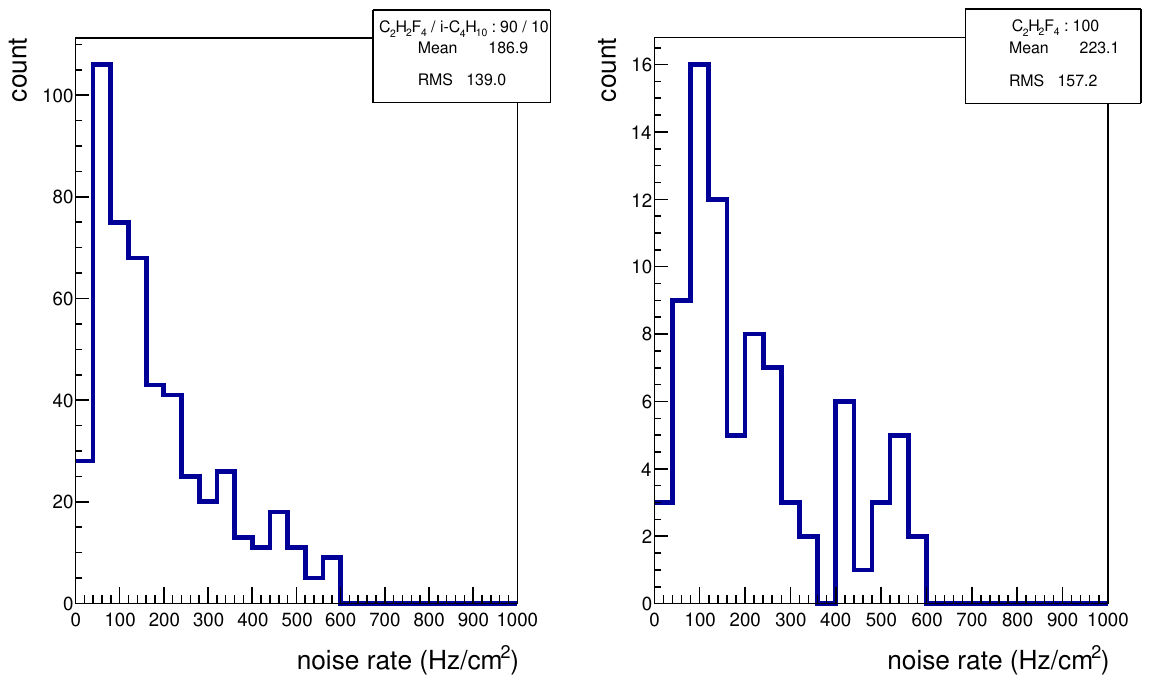}
	}
	\caption{Distribution of the noise rate values of long-term measurements at applied voltage of 10.2~kV.}
	\label{noise_distribution_long_term}
\end{figure}


The long-term stability of the detector is studied for about 100 days. During this measurement the ambient parameters such as  temperature, pressure and relative humidity are also recorded. The measured temperature (T) in Kelvin, pressure (p) in mbar and \% relative humidity (RH) as a function of the period of operation is shown in Figure~\ref{eff_parameter}. The mean T/p for the whole period is found to be 0.29~$\pm$~0.001~K/mbar. The measured efficiency and noise rate of the detector scaled for T/p at 0.29~K/mbar as a function of the period of operation for two different gas compositions are shown in Figure~\ref{eff_noise}. The distribution of efficiency and noise rate for two gas compositions are shown in Figure~\ref{eff_distribution_long_term} and \ref{noise_distribution_long_term} respectively. 

It is found that for C$_{2}$H$_{2}$F$_{4}$ and i-C$_{4}$H$_{10}$ mixture and 100\% C$_2$H$_2$F$_4$ the average efficiencies are found to be 89~$\pm$~6~\% and 94~$\pm$~6~\% respectively whereas the average noise rates for two compositions are found to be 187~$\pm$~139~Hz/cm$^{2}$ and 223~$\pm$~157~Hz/cm$^{2}$. Although the overall efficiency and noise rate for the entire period of $\sim$~100 days is found to be $\sim$~90\% and $\sim$~200~Hz/cm$^{2}$ respectively, but a large fluctuation is observed. One possible reason of this fluctuation is the following. The singles counting rate fluctuates largely because of the variation in the ambient parameters like temperature, pressure and relative humidity.

	\section{Summary and outlook}
	
	A RPC prototype made of linseed oil coated indigenous bakelite material of thickness 2~mm and having bulk resistivity $\sim$~3~$\times$~10$^{10}$~$\Omega$~cm is tested. The electrode material used here is bakelite high pressure paper laminates commercially available in the market.
	
	The detector is tested initially with 100\% C$_2$H$_2$F$_4$ gas. With the applied voltage of 10.6~kV, the time resolution obtained is $\sim$~1.03~$\pm$~0.03~ns. The time resolution of the module is also comparable with that of the conventional linseed oil-coated bakelite RPC.
			
	Variation of charge sharing between the consecutive strips is measured with the applied voltage. From 5~to 9~kV the shared charge fluctuates between 40-50\% after that it increases to $\sim$~60\%. With the increasing voltage shared charge has not increased further and remained constant with a slight fluctuations.
	
	The preliminary stability test of the chamber is performed using C$_{2}$H$_{2}$F$_{4}$ and i-C$_{4}$H$_{10}$ mixture and 100\% C$_2$H$_2$F$_4$ for about 100 days. The overall efficiency for the entire period is found to be $\sim$~90\%.
	
	This radiation tolerance test is very important for an experiment, especially for high-energy physics experiments where detectors are subjected to continuous operation at high rate environment for a long time. For the present prototype a very good efficiency is obtained where it is operated in the presence of a high-intensity photon source. It is also observed that the efficiency decreased by only 1~\% from the efficiency value without the source with a gamma ray flux of 46~kHz/cm$^{2}$.
	
	However, The operation mode of RPCs with pure Tetrafluoroethane (R134a) or with a binary mixture of 90\% Tetrafluoroethane and 10\% iso-butane is not the pure avalanche mode, as it was found in several crucial tests which were carried out between 1994 and 1998, both at the INFN laboratory in Rome and at CERN. At high voltage, the RPC signals in most events show an avalanche precursor followed by a much bigger streamer afterpulse. This urged researchers to find a way to suppress the streamer afterpulse, and the solution was found by adding a very small fraction of Sulphur Hexafluoride (SF$_6$$\sim$~0.3 \%), allowing 2-mm RPCs to operate in saturated avalanche mode with no streamer afterpulses. The chamber will be tested adding small fraction of SF$_6$ in future. A more detailed study on analog RPC signals with a wide-band oscilloscope, so that the integrated charge is measured for every efficient event, with the gas mixtures that is used to see events with only one peak (pure avalanche) and events with one avalanche signal followed by a bigger streamer afterpulse is also in future plan. With this information one can measure only the integrated avalanche charge in every event, and the integrated total charge, paying attention to subtracting the signal baseline which can fluctuate with respect to the zero-level of the oscilloscope due to external low-frequency noise.		
	\section{Acknowledgements}
	
	The authors would like to thank Prof. Sanjay K. Ghosh and Dr. Sidharth K. Prasad for their valuable discussions and suggestions during the course of the study. We would also like to thank Dr. Sumit Kumar Kundu, Indiana University, Bloomington, USA ; Mr. Ayan Dandapat of IIT Ropar, Punjab, India; Mr. Pranjal Barik of Savitribai Phule Pune University, Pune, India; and Mr. Ashwin Satheesan, Mahatma Gandhi University, Kerala, India for their help during data taking. Mr. Subrata Das is acknowledged for helping in the fabrication of the pick-up strips used in this study. This work is partially supported by the CBM-MuCh project from BI-IFCC, DST, Govt. of India. A. Sen acknowledges his Inspire Fellowship research grant [DST/INSPIRE Fellowship/2018/IF180361].


\begin{thebibliography}{99}
		
		\bibitem{santanico} R. Santonico and R. Cardarelli, Nucl. Inst. and Meth. A {\bf187} (1981) 377.

\bibitem{Biswas} S. Biswas {\it et al.}, Nucl. Inst. and Meth. A {\bf617} (2010) 138.

\bibitem{Park} S. K. Park {\it et al.}, 2012 JINST {\bf7} P11013.

\bibitem{Bhatt} A. D. Bhatt {\it et al.}, Nucl. Inst. and Meth. A {\bf844} (2017) 53.

\bibitem{BABAR} F. Anulli {\it et al.}, Nuclear Physics B (Proc. Suppl.) 61B (1998) 244.

\bibitem{ATLAS} The ATLAS collaboration {\it et al.} 2021 JINST 16 P07029.

\bibitem{ALICE_muon} A. Ferretti {\it et al.} 2019 JINST 14 C06011.

\bibitem{CMS} CMS - Technical Proposal, CERN/LHCC/94-38, December 1994.

\bibitem{ALICE_tof} M. Spegel {\it et al.} Nucl. Inst. and Meth. A {\bf453} (2000) 308.

\bibitem{STAR} E. Cerron Zeballos {\it et al.}, Nucl. Instr. and Meth. A {\bf374} (1996) 132.

 \bibitem{ARGO-YBJ} P Bernardini and (for the ARGO-YBJ Collaboration) 2008 J. Phys.: Conf. Ser. {\bf120} 062022.
 
\bibitem{COVER-PLASTEX} G.A. Agnetta {\it et al.}, Nucl. Instr. and Meth. A {\bf381} (1996) 64.

\bibitem{DAYABAY} Q. Zhang {\it et al.}, Nucl. Instr. and Meth. A {\bf583} (2007) 278.

\bibitem{CL09} C. Lu, Nucl. Instr. and Meth. A {\bf602} (2009) 761.

\bibitem{Abbrescia}M. Abbrescia {\it et al.}, Nucl. Instr. and Meth. A {\bf394} (1997) 13.

\bibitem{hong} B. Hong {\it et al.}, Journal of the Korean Physical Society, Vol. 48, No. 4, April 2006, 515.

\bibitem{babar_stalagmite}BaBar Technical Design Report, BaBar Collaboration, SLAC Report SLAC-R-95-457, March 1995.

\bibitem{Anulli_128} F. Anulli {\it et al.}, Nucl. Instr. and Meth. A {\bf508} (2003) 128.

\bibitem{Atlas} G. Cattani, Journal of Physics: Conference Series {\bf280} (2011) 012001.

\bibitem{cms} S. Park {\it et al.}, Nucl. Instr. and Meth. A {\bf550} (2005) 551

\bibitem{Zhang} J. Zhang {\it et al.}, Nucl. Instr. and Meth. A {\bf540} (2005) 102.

\bibitem{sb_2009}  S. Biswas {\it et al.}, Nucl. Instr. and Meth. A {\bf604} (2009) 310.

\bibitem{sen_2022} A. Sen {\it et al.}, Nucl. Instrum. Methods. A {\bf1024} (2022) 166095.

\bibitem{sen_2022_2} A. Sen {\it et al.}, Nucl. Instrum. Methods. A {\bf1045} (2023) 167572.

\bibitem{PZ2012} Peng Zhang, PhD thesis, University of Michigan, 2012, UMI Number: {\bf 3554236}.
	
\bibitem{SPM1949} S. P. Morgan, J. Appl. Phys. 20, {\bf352} (1949).

\bibitem{SG2017} S. Groiss, {\it et al.}, IEEE Trans. Magnetics 32, {\bf894} (1996).

\bibitem{ME1971} Microwave Engineers Handbook, Vol. Two (Artech House, 1971) p.{\bf186}.

\bibitem{sahu} S. Sahu {\it et al.}, JINST {\bf12} (2017) C05006.

\bibitem{sen_2020} A. Sen {\it et al.}, JINST {\bf15} (2020) C06055.
		
		
		
		
	\end{thebibliography}
\end{document}